\documentstyle[prl,aps]{revtex}

\begin{document}

\draft

\title{Response to Parisi's Comment
on ``Non-mean-field behavior of realistic spin glasses''}  
\author{C.M. Newman}
\address{Courant Inst. of Math. Sciences,
New York Univ., New York, NY 10012
and Inst. for Adv. Study, Princeton, NJ}
\author{D. L. Stein}
\address{Department of Physics, University of Arizona,
Tucson, AZ 85721}
\maketitle

\begin{abstract} 
We expand on why our recent results rule out the
standard SK picture of realistic spin glasses.
\end{abstract}

\narrowtext

In his Comment \cite{Parisi} on our paper \cite{NS96}, 
Parisi argues that our two constructions
of $P_J$ are self-averaging only 
because neither is the ``correct'' one, 
which is an infinite volume object (i.e., $L$ has already been taken to
$\infty$) that {\it does} depend on $J$ (in the spin glass phase). 

We discuss below Parisi's analysis of our constructions.
But first, we make a more important
point, concerning the conventional formulation
(as in \cite{Parisi}) of the mean field
predictions for realistic models (called the ``standard SK picture''
in \cite{NSpreprint}):  a central conclusion of
\cite{NS96} is that there
cannot exist {\it any} $P_J$ which is both an
infinite volume object {\it and\/} which depends on $J$ --- 
at least not unless
it has the physically peculiar property of depending on the choice of the origin
of the coordinate system.  This is an immediate (and rigorous) consequence
of the spatial ergodicity of the underlying disorder distribution, as
explained in \cite{NS96}:  any translation invariant (infinite volume) $P_J$ must be
self-averaging. The standard SK picture is therefore self-contradicting.    

What about the claims in \cite{Parisi} about our two constructions?
We first note that the two constructions of \cite{Parisi} are not
the same as ours, because the latter
implicitly take an overlap $q_{R'}$ with $R'\ne R$,
the box size where the couplings are fixed.  For example,
for the second $P_J$ of \cite{NS96}, which we denote by
$P_J^{II}$, $R'\ll R$. These differences are subtle, but can lead to quite
different $P_J$'s due to finite size effects, as emphasized in \cite{HF}.
We will not dwell on these issues here, but will address Parisi's
claim that our second $P_J(q)$ is independent of $J$
because it is a delta-function. 

{\bf 1.}  Must $P_J^{II}(q) = \delta(q)$ 
due to the ``chaotic nature of spin glasses'', as
asserted in \cite{Parisi}?  No;  
the ``nonstandard'' SK picture of
\cite{NSpreprint} has just such a chaotic nature,
but there $P_J^{II}(q)$ could be continuous and nonzero everywhere
between $\pm q_{EA}$ (with no delta-functions at those points).

{\bf 2.}  Could $P_J^{II}(q) = \delta(q)$ when there are many pure states? 
Yes; as already noted in \cite{NS96}, this
occurs in the model of Ref.~\cite{NS94}. For realistic models,
it could occur in the context of
possibility 5 in Ref.~\cite{NSpreprint}
and would mean that
the overlaps between pure states are not a good choice of order parameter.

In \cite{Parisi}, Parisi defines
his $P_J(q)$ first for finite L and then 
takes $L\to\infty$.  But
if there are many pure states, then this limit
should not exist because of chaotic size dependence \cite{NS92}. 
We have not found
in the literature any construction
(other than ours or related ones \cite{AW}) of 
a natural infinite-volume $P_J(q)$
for short-ranged spin glasses.  We would welcome
such a construction,
but we emphasize that
any infinite-volume $P_J(q)$ which has the very weak and natural property
of translation-invariance will be automatically self-averaging.

Does all this prove that mean-field theory is 
irrelevant to realistic spin glasses?
Not yet. In \cite{NSpreprint}, we present an approach to 
realistic disordered (and other) systems,
which {\it might\/} allow some mean-field features
to persist.  A key
aspect of this approach is that in infinite volume, dependence on $J$ is
replaced by a more subtle type of dependence.  As discussed 
in \cite{NSpreprint},
this type of dependence is fully
consistent both with the observation of 
Guerra \cite{Guerra} that taking replicas and
infinite volume limits in different orders could lead to different results,
and with the possibility of replica symmetry breaking.
However, the resulting nonstandard SK picture
differs considerably from the standard
one; in particular, there is no 
dependence of (infinite volume) overlap distributions
on $J$ and there cannot be ultrametricity of
overlaps among {\it all\/} pure states.  We refer the reader
to \cite{NSpreprint} for further details.

\acknowledgments

This research was partially supported by NSF Grant 
DMS-95-00868 (CMN) and by DOE Grant DE-FG03-93ER25155 (DLS).



\begin{references}

\bibitem {Parisi} G.~Parisi, 
\newblock  ``Recent rigorous results 
support the predictions of spontaneously broken
replica symmetry for realistic spin glasses'', preprint, March, 1996.
Available as cond-mat preprint 9603101 at http://www.sissa.it/.
 
\bibitem {NS96}  C.M.~Newman and D.L.~Stein, 
\newblock {\em  Phys.~Rev.~Lett.\/} {\bf 76}, 515 (1996).

\bibitem {NSpreprint}  C.M.~Newman and D.L.~Stein, 
\newblock ``Spatial inhomogeneity and thermodynamic chaos'',
October 1995.  Available at http://www.physics.arizona.edu/~dls, 
or as adap-org preprint \# 9511001. 

\bibitem {HF}  D.A.~Huse and D.S.~Fisher,
\newblock {\em J.~Phys.~A\/} {\bf 20}, L997 (1987).

\bibitem {NS94}  C.M.~Newman and D.L.~Stein, 
\newblock {\em  Phys.~Rev.~Lett.\/} {\bf 72}, 2286 (1994).

\bibitem {NS92}  C.M.~Newman and D.L.~Stein, 
\newblock {\em Phys.~Rev.~B} {\bf 46}, 973 (1992). 

\bibitem{AW}
M.~Aizenman and J.~Wehr,
\newblock {\em Comm.~Math.~Phys.\/} {\bf 130}, 489 (1990);
A.~Gandolfi, M.~Keane, and C.M.~Newman,
\newblock {\em Prob.~Theory~Rel.~Fields\/} {\bf 92}, 511 (1992).

\bibitem{Guerra}
F.~Guerra, private communication; see also
``About the overlap distribution in mean
field spin glass models'', preprint, November, 1995.






\end{references}
\end{document}